\documentclass[runningheads]{llncs}
\usepackage[T1]{fontenc}
\usepackage{graphicx}
\usepackage[hidelinks]{hyperref}
\usepackage{color}

\usepackage{multirow}
\usepackage{mdframed}
\usepackage{graphicx}
\usepackage{tabularx}
\usepackage{xltabular}
\usepackage{svg}
\usepackage{placeins}
\usepackage{float}
\usepackage{todonotes}
\usepackage{amssymb}

\usepackage{longtable}
\usepackage{booktabs}
\usepackage{array}

\newcommand*{\fullref}[1]{\hyperref[{#1}]{\autoref*{#1} \nameref*{#1}}}
\newcommand*{\secref}[1]{\hyperref[{#1}]{\autoref*{#1}}}
\begin{document}
\title{ExtracTable: Human-in-the-Loop Transformation of Scientific Corpora into Structured Knowledge}
\titlerunning{ExtracTable: From Corpora to Structured Knowledge}
%
\author{Lena John\inst{1}\orcidID{0009-0007-2097-9761} \and Ahmed Malek Ghanmi\inst{2} \and
Tim Wittenborg\inst{3,4}\orcidID{0009-0000-9933-8922} \and
Sören Auer\inst{1,3,4}\orcidID{0000-0002-0698-2864} \and Oliver Karras\inst{1}\orcidID{0000-0001-5336-6899} }
\authorrunning{L. John et al.}
%
\institute{TIB - Leibniz Information Centre for Science and Technology, Hannover, Germany\\
\email{\{lena.john, soeren.auer,  oliver.karras\}@tib.eu}
\and
Leibniz University Hannover, Hannover, Germany 
\and
L3S Research Center, Leibniz University Hannover, Hannover, Germany\\
\email{\{tim.wittenborg, soeren.auer\}@l3s.de} \and Cluster of Excellence SE²A – Sustainable and Energy-Efficient Aviation, Technische Universität
Braunschweig, Germany}
\maketitle              
\begin{abstract}
As the volume of scientific literature grows, efficient knowledge organization becomes increasingly challenging. Traditional ap\-proach\-es to structuring scientific content are time-consuming and require significant domain expertise, highlighting the need for tool support.
We present \textit{ExtracTable}, a Human-in-the-Loop (HITL) workflow and framework that assists researchers in transforming unstructured publications into structured representations. The workflow combines large language models (LLMs) with user-defined schemas and is designed for downstream integration into knowledge graphs (KGs). Developed and evaluated in the context of the Open Research Knowledge Graph (ORKG), \textit{ExtracTable} automates key steps such as document preprocessing and data extraction while ensuring user oversight through validation.
In an evaluation with ORKG community participants following the Quality Improvement Paradigm (QIP), \textit{ExtracTable} demonstrated high usability and practical value. Participants gave it an average System Usability Scale (SUS) score of 84.17 (A+, the highest rating). The time to progress from a research interest to literature-based insights was reduced from between 4 hours and 2 weeks to an average of 24:40 minutes.
By streamlining corpus creation and structured data extraction for knowledge graph integration, \textit{ExtracTable} leverages LLMs and user models to accelerate literature reviews. However, human validation remains essential to ensure quality, and future work will address improving extraction accuracy and entity linking to existing knowledge resources.

\end{abstract}
\section{Introduction}
Conducting literature reviews is indispensable for scientific research.
This task is inherently challenging, as it requires a significant amount of time and effort to gain an overview of the existing research and knowledge on a specific topic, helping to support methodology, contextualize research, or identify gaps in knowledge~\cite{snyder_literature_2019}.
With the rapid growth in scientific discovery and distribution of knowledge, it has become even more difficult to produce high-quality reviews~\cite{larsen_understanding_2019}.
Recent advancements in artificial intelligence (AI), particularly the rise of large language models (LLMs) and AI-based Literature Reviews~\cite{wagner_artificial_2022}, offer promising opportunities to streamline and automate repeatable and time-consuming processes~\cite{rathi_p21_2023}.
However, despite their growing popularity, AI models face significant limitations, primarily the tendency to produce hallucinations, resulting in inconsistencies and confusion~\cite{perkovic_hallucinations_2024}.
Combining AI capabilities with a consistent knowledge base, called neuro-symbolic AI, mitigates these risks by leveraging reliable knowledge representations to alleviate AI's unreliability~\cite{Auer.2025}.
The knowledge bases need to be Findable, Accessible, Interoperable, and Reusable (FAIR)~\cite{wilkinson_fair_2016} to address the requirements of the scholarly domain and AI technology~\cite{Auer.2023}.
Such solutions already exist, like the Open Research Knowledge Graph (ORKG)\footnote{\url{https://orkg.org/}} and its neuro-symbolic service Ask~\cite{Auer.2025}, offering a novel approach by representing scientific information in a structured format, providing both human- and machine-readability.
However, organizing knowledge at scale reintroduces the original challenge of requiring human oversight, leading to manual work and a circular dependency.
To address this issue, an iterative, Human-in-the-Loop (HITL), AI-assisted knowledge organization approach is needed, constantly improving scholarly knowledge representation.
Our contributions include
1) \textit{ExtracTable}, a Human-in-the-Loop (HITL) approach comprising a workflow and modular implementation that combines large language model (LLM) support with Natural Language Processing (NLP) techniques to streamline, automate, and formalize literature review tasks.
2) The modular implementation of \textit{ExtracTable} as a Jupyter-based framework to build scientific corpora and transform unstructured information into structured knowledge ready for knowledge graph integration.
3) An iterative evaluation with nine researchers over four months, demonstrating \textit{ExtracTable}’s ability to improve efficiency while maintaining user oversight of automated processes.
This work is 
structured as follows:
In \secref{sec:background} we give an overview of the background and related work in the HITL context. Our approach is described in \secref{sec:approach}, followed by an outline of the implementation in \secref{sec:implementation}.
We evaluate the developed solution of \textit{ExtracTable} in \secref{sec:evaluation} and discuss the results in \secref{sec:discussion}. 
We conclude our work in \secref{sec:conclusion}.

\section{Background and Related Work\label{sec:background}}
This section explores Knowledge Graphs (KGs), AI-driven automation, and HITL systems in scholarly knowledge organization, highlighting their roles in literature reviews and knowledge integration.\vspace{0.1cm}

\noindent
\textit{Knowledge Organization.\label{sec:background_knowledge}}
Knowledge Graphs (KGs) are designed to represent and interconnect information in a structured, machine-readable format~\cite{vahdati_unveiling_2018}, like the scientific knowledge graph (SKG) Semantic Scholar~\cite{Wade2022TheSemanticScholar}.
Manually preparing data for KGs, particularly SKGs, is time-consuming and lacks scalability, as it requires domain experts to search, interpret, and structure scientific literature~\cite{hofer_construction_2024}.
Without domain experts, there is a higher risk of integrating incorrect, incomplete, or outdated information into the SKG, reducing its reliability and quality~\cite{jain_domain-specific_2020}.
The quality of SKGs heavily depends on their construction method, with significantly different results between manual and automated methods~\cite{huaman_knowledge_2022}.
The ORKG is a versatile scientific knowledge platform that uses crowdsourcing to organize and link scientific knowledge across various topics and domains~\cite{Auer.2025}.
Karras et al.~\cite{karras_divide_2023,karras_kg-empire_2024,Karras2025,Karras.2024} demonstrated how the ORKG can enhance and ensure FAIR~\cite{wilkinson_fair_2016} literature reviews, highlighting their importance in providing research with structured, long-term available data.
The ORKG Assistant for Scientific Knowledge (Ask)~\cite{Auer.2025} applies a neuro-symbolic approach using semantic search, LLMs, and KGs.
It exemplifies a next-generation scholarly search and exploration system, combining semantic knowledge representation with AI.\vspace{0.1cm}

\noindent
\textit{Artificial Intelligence (AI).}
AI and NLP techniques can automate content generation and assignment~\cite{dessi_generating_2021}, but these methods often lack transparency and accountability while struggling with semantic context~\cite{khalid_explainable_2022}.
Fully relying on AI-supported technologies poses significant risks, making human supervision and validation indispensable~\cite{ren_human-machine_2023}.
The study conducted by Danler et al.~\cite{danler_quality_2024} demonstrates that, while AI-generated outputs can be of high quality, ``\textit{relying solely on AI-generated outputs for crafting a scientific paper is not advisable}''~\cite[p.~207]{danler_quality_2024}.
Bacinger et al.~\cite{bacinger_system_2022} present Litre Assistant, a machine-learning-based solution to determining a corpus of scientific papers as the initial task of the Systematic Literature Review (SLR) process.
Their work introduces a system that allows users to iteratively select desired papers as a set of training data for machine learning algorithms for an automated classification process.
Ye et al.~\cite{ye_hybrid_2024} propose a hybrid, ``human-oriented'', but AI-based workflow to support specific steps of an SLR.
Their workflow screens full-text articles to extract relevant data and allows the user to perform inclusion and exclusion decisions, resulting in a spreadsheet of LLM-generated data.
The work of Alshami et al.~\cite{alshami_harnessing_2023} introduces a method where humans review ChatGPT-generated data for literature setup, abstract screening, data extraction, and analysis. However, relying solely on ChatGPT restricts access to real-time data, limiting the research scope to the model's training data and reducing its applicability.\vspace{0.1cm}

\noindent
\textit{Human-in-the-Loop (HITL).}
Several previously mentioned works incorporate human oversight in selection, classification, processing, and representation to achieve more accurate results~\cite{alshami_harnessing_2023,Auer.2025,bacinger_system_2022,ye_hybrid_2024}.
Consequently, research increasingly recognizes the value of HITL approaches~\cite{kommineni_human_2024}.
Through providing iterative feedback, users validate outputs as well as refine and enhance the performance system over time to bridge the gap between efficiency and contextual relevance~\cite{dalvi_mishra_towards_2022}.
Schatz et al.~\cite{schatz_workflow_2022} developed a workflow for curating a biomedical KG (DRKG)~\cite{ioannidis_drkg_2020}, that integrates HITL components. 
Their approach is particularly relevant as it addresses HITL usability aspects in the context of curating and enriching KGs.
The work by Tsavena et al.~\cite{tsaneva_enhancing_2024} builds on the automated pipeline for curating SKGs introduced in SCICERO~\cite{dessi_scicero_2022}.
They propose a HITL validation module as a solution to incorporate human decision-making into the curation process.
This module, one of two proposed validation mechanisms, establishes a hybrid approach that integrates human expertise to enhance the curation process.
The authors report a significant improvement in the accuracy of automated KG generation, demonstrating an improvement in both precision and recall.
SWARM-SLR by Wittenborg et al.~\cite{wittenborg_swarm-slr_2024} establishes a literature review workflow implemented by a modular framework that utilizes several AI-based tools to fulfill 65 requirements derived from SLR guidelines.
While many requirements can be supported or even fulfilled by AI, the overall workflow still requires human expert intervention at several stages of the literature review, particularly for knowledge extraction and validation.
Users remain crucial for guiding and validating tasks where automated approaches lack accuracy or contextual understanding~\cite{mosqueira-rey_human---loop_2023}.

Our work builds on the discussed approaches, emphasizing the role of HITL in the context of AI-assisted knowledge graph construction.
In this spirit, we aim to develop a solution that transforms unstructured input into structured, human-validated outputs designed for downstream knowledge graph integration. Once integrated into knowledge graphs, these structured outputs can support the human- and machine-readable context needed by neuro-symbolic AI to sustainably advance scholarly knowledge organization.

\section{Approach\label{sec:approach}}
We adopt the Quality Improvement Paradigm (QIP)~\cite{basili_experience_2002} to guide the iterative development of \textit{ExtracTable}, a Human-in-the-Loop (HITL) approach that includes a workflow and is realized as a modular framework for transforming scientific corpora into structured knowledge.

\begin{figure}[h!]
  \centering
    {
        \includegraphics[width=0.8\textwidth]{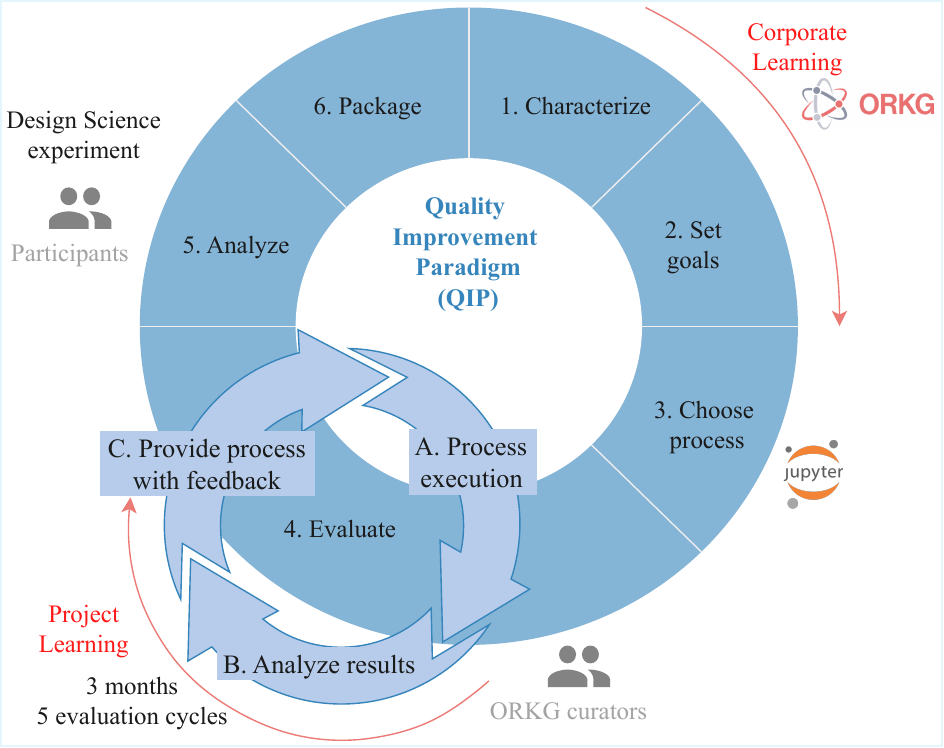}
  \caption{Application of the Quality Improvement Paradigm (QIP) methodology to develop \textit{ExtracTable}, together with ORKG curators. Derived from~\cite{oliveira_software_2014}.}}
  \vspace{-5mm}
  \label{fig:qip}
\end{figure}

\textit{ExtracTable} was developed in close collaboration with ORKG curators, researchers from diverse domains funded through grants to create structured comparisons, a key content type presenting tabular overviews of state-of-the-art research in the ORKG~\cite{Oelen.2019}. While their financial affiliation may introduce some bias, their interest in improving tools that support their contributions fosters practical, needs-driven feedback.
An overview of the development process is shown in~\secref{fig:qip}, with detailed tasks in~\secref{tab:qip_steps}. The \textit{Project Learning} cycle focused on short-term usability improvements based on curator feedback, while the \textit{Corporate Learning} cycle addressed long-term enhancements to ORKG infrastructure beyond the project.

\begin{longtable}{p{0.25\textwidth} p{0.75\textwidth}}
\caption{QIP step descriptions for the development of \textit{ExtracTable}} \label{tab:qip_steps} \\
\toprule
\textbf{Step} & \textbf{Description} \\
\midrule
\endfirsthead
\endhead

\endfoot
\textit{1. Characterize} & In the first phase, we examined research workflows, including topic selection, corpus retrieval, and structured overview creation. Collaborating with ORKG curators, we identified key content and technical requirements, resulting in a modular, adaptable framework for diverse SKG use cases.\\\hline

\textit{2. Set goals} & We defined functional and non-functional requirements to set the tool’s scope, ensuring it addresses key pain points like time-consuming corpus creation and limited support for structured output (cf.~\secref{sec:requirements}).\\\hline

\textit{3. Choose process} & We selected external ORKG curators as a focus group due to their research diversity and active involvement in literature-based knowledge graph curation. Using an iterative participatory design approach with bi-weekly meetings, we co-developed modular Jupyter Notebooks for \textit{ExtracTable} tailored to flexible workflow stages (cf.~\secref{sec:approach_workflow}).\\\hline

\textit{4. Execute} & In this phase, we implemented and iteratively refined the framework with ongoing feedback. Over three months in five meetings, we made key design choices, including abstract-only publication support and splitting the workflow into two notebooks (cf.~\secref{sec:implementation}). \\\hline

\textit{5. Analyze} & We finalized the framework by incorporating feedback and evaluating its effectiveness through a design science experiment. This revealed usability issues, guided feature refinements, and validated alignment with real-world curation needs (cf.~\secref{sec:evaluation}). \\\hline

\textit{6. Package} & Finally, we compiled reusable artifacts for \textit{ExtracTable}, including framework components, evaluation metrics, and key insights, to support broader adoption. We also assessed ORKG’s infrastructure to enable sustainable integration (cf. sections~\ref{subsec:results} and~\ref{sec:discussion}).\\
\end{longtable}

\subsection{Requirements}\label{sec:requirements}
To define functional and non-functional requirements for \textit{ExtracTable} (see~\secref{tab:requirements}), we drew inspiration from Wittenborg et al.~\cite{wittenborg_swarm-slr_2024}, who specified requirements for the typical phases of an SLR. While our framework does not aim to replicate or enforce the rigor of a formal SLR, we adopt this process-oriented perspective to ensure relevance and alignment with real-world research workflows. In collaboration with the SWARM-SLR curators and domain experts, we adapted these ideas to guide the design of system requirements and support a more lightweight, customizable process for literature exploration. Our goal is to offer a modular framework for \textit{ExtracTable} that enables generation and processing of a literature corpus without the complexity of a full SLR pipeline. To ensure modularity and usability, the workflow is divided into two stages, and users can start with the second if a relevant corpus already exists.

\begin{longtable}{m{0.85\textwidth}|>{\centering\arraybackslash}m{2cm}}
    \caption{Functional (FR) and non-functional (NR) requirements}\label{tab:requirements} \\
    \textbf{Functional Requirements (FR)} & \textbf{Fulfillment} \\
    \hline
    \endfirsthead
    \hline
    \endhead
    \endfoot
    \endlastfoot
    FR1: The framework should suggest relevant keywords based on a user-defined research interest. & Yes\\\hline
    FR2: The framework must allow a user to manually input and modify a search query to retrieve relevant literature. & Yes\\\hline
    FR3: The framework must support integration with external literature databases. & Yes\\\hline
    FR4: The framework must provide options to configure search parameters. & Yes\\\hline
    FR5: The framework should present retrieved search results in an interactive format. & Yes\\\hline
    FR6: The framework must enable a user to select and download relevant papers to create a literature corpus. & Yes\\\hline
    FR7: The framework must allow a user to use an existing corpus to proceed with knowledge extraction. & Yes\\\hline
    FR8: The framework must provide an interface to define a customizable data model, specifying the properties to extract from each publication. & Partial\\\hline
    FR9: The framework must extract information according to the user-defined data model from the provided literature corpus. & Yes\\\hline
    FR10: The framework should flag properties that cannot be extracted or are missing to alert users. & Yes\\\hline
    FR11: The framework must present the extracted data in an editable format to allow user refinement and approval. & Yes\\\hline
    FR12: The framework will support entity linking on the extracted and validated content for seamless KG-integration. & No\\\hline
    \textbf{Non-Functional Requirements (NR)} & \textbf{Fulfillment} \\\hline
    NR1: The framework must be implemented in a modular way, supporting the integration of new data sources or processing steps. & Yes\\\hline
    NR2: The framework should offer an interactive user interface. & Yes\\\hline
    NR3: The framework must unify data returned from various data sources into a consistent internal schema to support comparison. & Yes\\\hline
    NR4: The framework should be capable of handling literature corpora of varying sizes without significant loss in performance. & No\\\hline
    NR5: The framework should be well-documented, follow clean code principles, and be openly available to support long-term maintenance, reproducibility, and community contribution. & Yes\\\hline
\end{longtable}

\subsection{Workflow\label{sec:approach_workflow}}
We now describe how the \textit{ExtracTable} approach, centered on its HITL workflow, meets the requirements outlined above. While not aiming to replicate the full SLR process, the two workflow stages draw inspiration from its typical phases and provide a practical structure for key tasks. Our workflow blends automation with human validation, enabling users to focus their expertise on interpreting and refining scientific content. To reduce manual effort, we automate time-consuming steps while preserving user control over key decisions. The workflow can start from any research interest without requiring pre-structured input, making it flexible and adaptable to diverse review scenarios.\vspace{0.2cm}

\begin{centering}
\begin{figure}[h!]
    \vspace{-0.4cm}
    {
    \fontsize{4}{6}
    \centering
    \includegraphics[width=1.1\columnwidth]{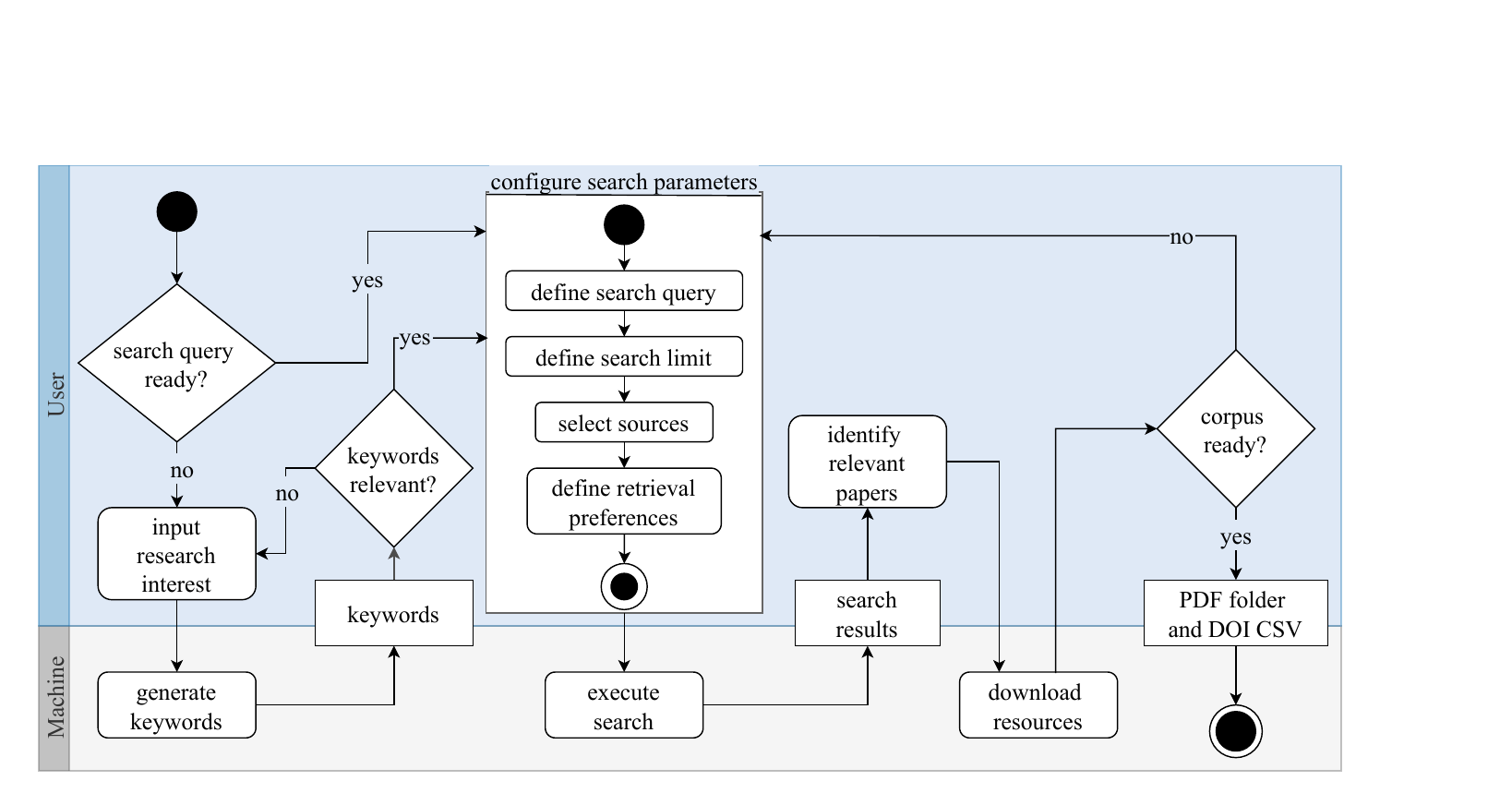}
    }
    \vspace{-0.4cm}
    \caption{Activity diagram for \textit{Stage 1: Creating a scientific corpus of literature}, demonstrating user-machine interaction in the scientific literature corpus building process.}
    \vspace{-0.4cm}
  \label{fig:activity1}
  \vspace{-0.3cm}
\end{figure}
\end{centering}

\noindent
\textit{Stage 1: Creating a scientific corpus of literature.}
This stage focuses on building a corpus of publications relevant to the user’s research interest (see~\secref{fig:activity1}). After defining a research interest and optionally using the LLM to suggest keywords, users formulate a search query to retrieve publications from selected sources. Search parameters, such as output size or restricting to openly accessible full texts, are configurable to suit user preferences. Retrieved papers are reviewed and selected by the user, forming the input for Stage 2.\vspace{0.2cm}

\begin{figure}[H]
\vspace{-0.3cm}
  \centering
    \fontsize{4}{6}
    \includegraphics[width=\textwidth]{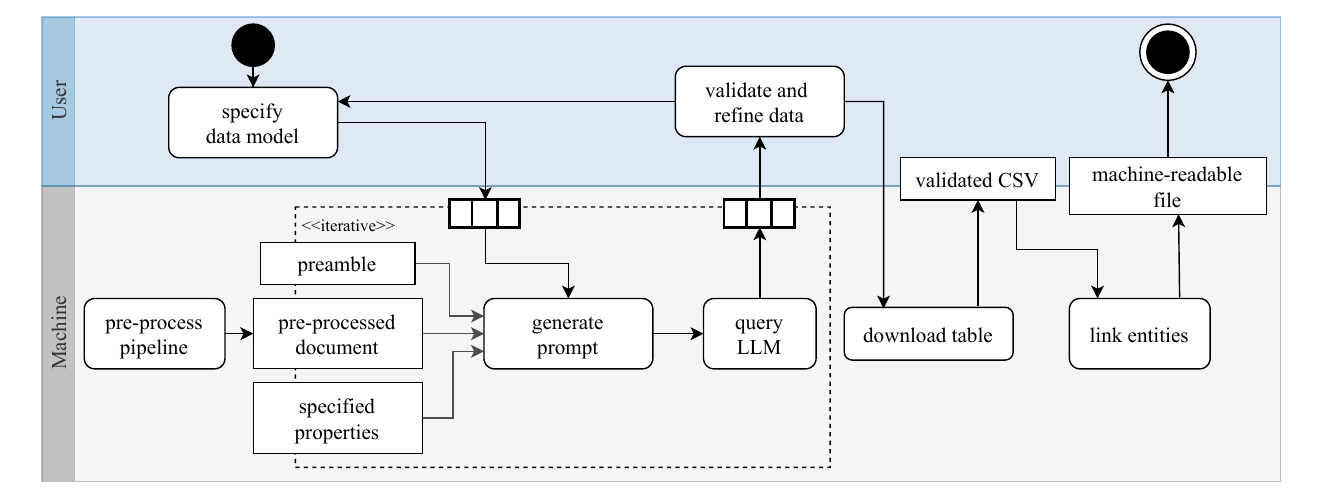}
    \vspace{-0.4cm}
    \caption{Activity diagram for \textit{Stage 2: Knowledge extraction and validation}, showing the tasks executed by both the user and machine to extract knowledge from the corpus.}
    \vspace{-0.4cm}
    \label{fig:activity2}
\end{figure}

\noindent
\textit{Stage 2: Knowledge extraction and validation.}
The second stage processes the knowledge within the corpus, as shown in~\secref{fig:activity2}. The user begins by defining the properties of the data model to be extracted from each paper. The LLM agent then parses and interprets the corpus, generating output based on the specified properties. If any information is not found, the LLM flags this to mitigate the risk of hallucination. The user can manually refine, revise, and modify the results to ensure accuracy, ultimately validating the output. The final results are formatted, and optionally, entity linking is performed to create output that is suitable for KG integration as a machine-readable file.

\section{Implementation\label{sec:implementation}}
We implement the described functions as a modular framework for \textit{ExtracTable} in two \href{https://gitlab.com/TIBHannover/orkg/ExtracTable}{\textcolor{blue}{Jupyter Notebooks}} using Python (NR5). These notebooks provide automated modules, visualizations, and browser-rendered interactive controls (NR2). To enhance usability, we leverage interactive HTML widgets from \href{https://pypi.org/project/ipywidgets/}{\textcolor{blue}{ipywidgets}} and the \href{https://ipython.org/}{\textcolor{blue}{IPython}} toolkit for abstracting complex code interactions. Users can configure workflows via UI elements such as input fields and checkboxes, and extend functionality by adding query sources or modifying the UI and behavior as needed (NR1). The implementation is modular and flexible, designed to support integration with various knowledge graphs.

\begin{figure}[H]
\vspace{-0.4cm}
    \centering
    \includegraphics[width=1.0\textwidth]{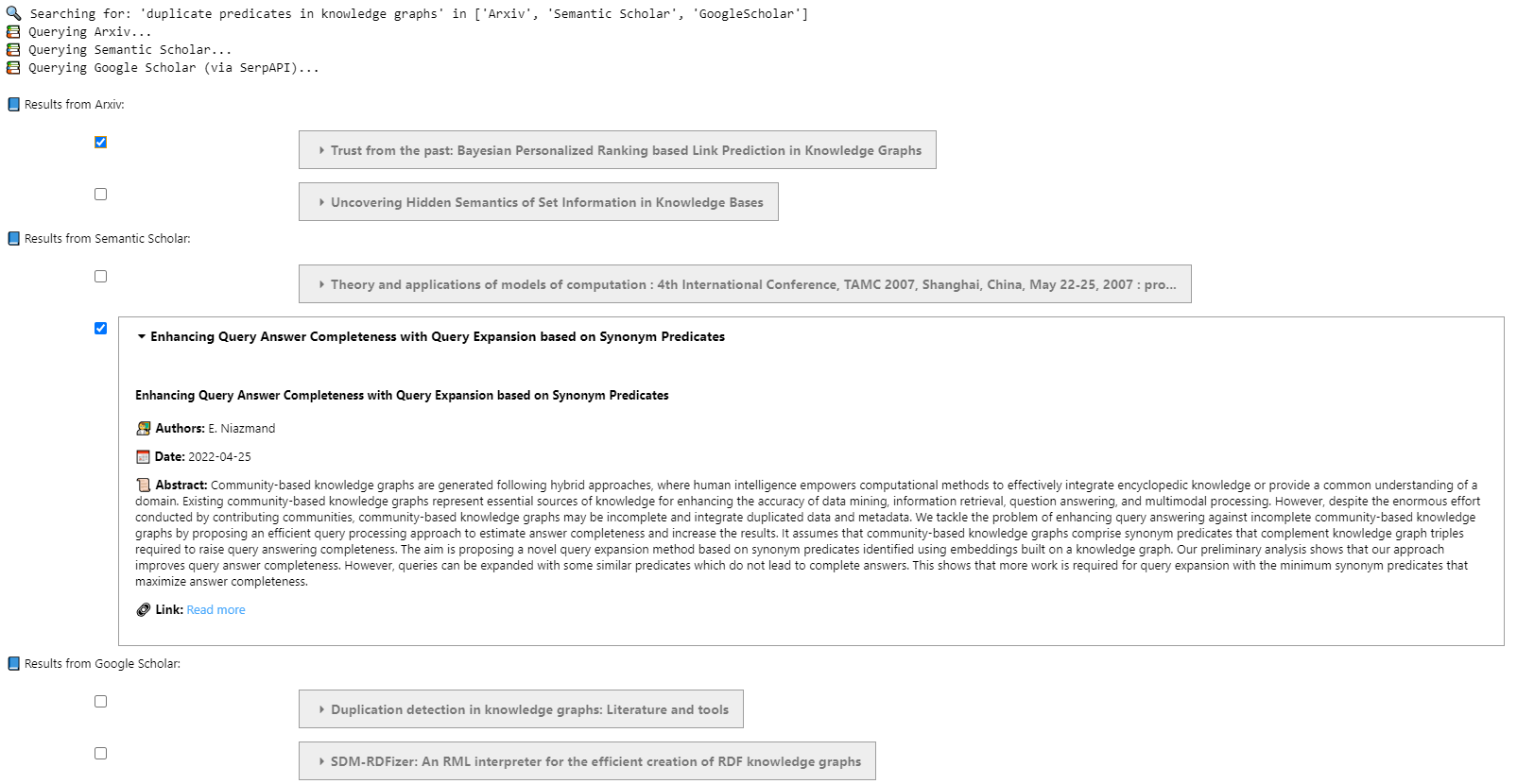}
    \caption{Visualization of query results, where each paper is presented within an expandable accordion, providing access to detailed information relevant to the paper.}
    \vspace{-0.4cm}
    \label{fig:fig_task_stage1}
\end{figure}

The \textbf{first stage} retrieves relevant papers based on a user-defined search query (FR2). To support query formulation, the framework can optionally suggest keywords from the user’s research interest (FR1). We integrate multiple APIs, including \href{https://www.semanticscholar.org/product/api}{\textcolor{blue}{SemanticScholar}}, \href{https://info.arxiv.org/help/api/index.html}{\textcolor{blue}{ArXiv}}, and the \href{https://pypi.org/project/serpapi}{\textcolor{blue}{SerpApi wrapper}}, enabling real-time searches across diverse scholarly databases (FR3) based on user preferences (FR4). Returned JSON objects with varying schemas are unified for consistency (NR3). Key elements such as title, abstract, authors, publication date, and full-text links are presented within expandable accordions in the notebook (FR5), as shown in~\secref{fig:fig_task_stage1}. Alternatively, users may work with an existing corpus, which can currently be imported from a Zotero collection (FR7). 

Based on these suggestions, the user iteratively creates the corpus by selecting and downloading relevant papers (FR6). For reproducibility, the corresponding DOIs can be exported to a CSV file.

\begin{figure}[H]
    \vspace{-0.4cm}
    \centering
    \includegraphics[width=1.0\textwidth]{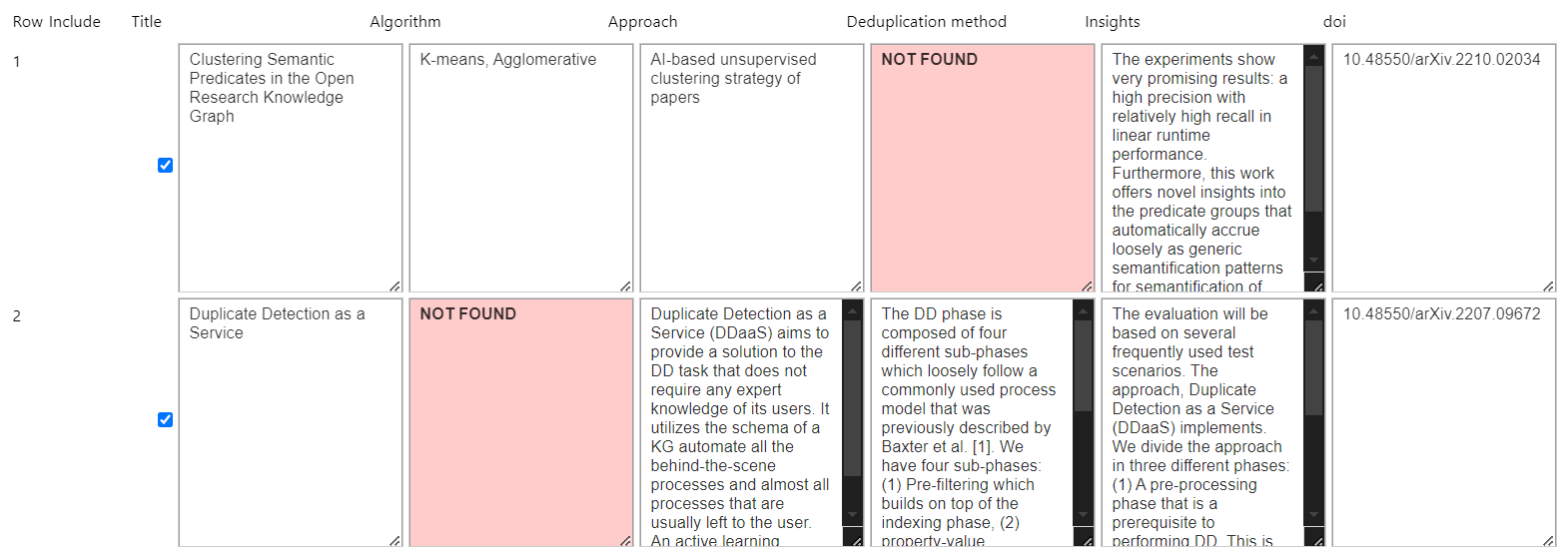}
    \vspace{-0.2cm}
    \caption{Visualization of LLM-generated information. If a property is not found in the text, the tool renders a label ``NOT FOUND'' and displays it in red (FR10).}
    \vspace{-0.4cm}
    \label{fig:datagrid_stage2}
\end{figure}

The \textbf{second stage} processes the corpus using an LLM. We use the \href{https://mistral.ai/news/mistral-large-2407/}{\textcolor{blue}{Mistral-large-2 model}}, accessible via Mistral’s API and currently free of charge. This model supports JSON-formatted outputs, enhancing structure and compatibility with downstream workflows.
The tool prompts the LLM iteratively using user-defined properties (FR8) and pre-processed text extracted from each PDF. Pre-processing starts with raw text extraction via \href{https://pypi.org/project/pdfminer/}{\textcolor{blue}{pdfminer}}. References and appendices are removed, and regex heuristics clean and reconstruct sentences and paragraphs to preserve context.
Generated outputs (FR9) are displayed in a datagrid (FR11; see~\secref{fig:datagrid_stage2}), allowing users to review, edit, and select entries for final CSV export. Users can also refine the data model and rerun extraction, though previous results may be overwritten. An example of prompt and output can be found in the \href{https://gitlab.com/TIBHannover/orkg/ExtracTable/-/blob/main/example.md}{\textcolor{blue}{repository}}.

We have begun integrating entity linking (FR12) as a step toward semantic enrichment. At this stage, we use \href{https://labs.tib.eu/falcon/falcon2/}{\textcolor{blue}{Falcon 2.0}}\cite{sakor_falcon_2020} for entity extraction, identifying relevant concepts in free-text fields. Although Falcon already supports linking entities to knowledge graphs such as DBpedia\cite{auer_dbpedia_2007} and Wikidata~\cite{vrandecic_wikidata_2014}, we currently focus on extraction as a first step toward machine-actionability. This approach prepares the ground for entity linking and, ultimately, seamless integration with ORKG and other KGs in the future. Previous work~\cite{John_Scimantify.2025} examined how knowledge representation evolves across formats and focused on semantically enriching tables by aligning entities and structuring content during KG import.

\section{Evaluation\label{sec:evaluation}}
The framework \textit{ExtracTable} is evaluated through a design science experiment involving prospective users to assess its practical usability and real-world applicability. To contextualize this evaluation, we consider the ORKG as a representative environment for scientific content creation. Following the goal definition template by Wohlin et al.~\cite{wohlin_experimentation_2012}, we define the experiment’s objective as follows:

\begin{mdframed}
\textit{Research goal:} \textbf{Analyze} \textit{ExtracTable}
\textbf{with the purpose of} supporting systematic content creation for scientific KGs
\textbf{in terms of} usability (effectiveness, efficiency, satisfaction)
\textbf{from the perspective of} potential future users familiar with the ORKG
\textbf{in the context of} the ORKG.
\end{mdframed}
The goal leads to the research question:
\begin{mdframed}
    \textit{Research question:} How does the developed tool \textit{ExtracTable} impact the usability of systematic content creation for scientific knowledge graphs, from the perspective of potential future users familiar with the ORKG?
\end{mdframed}
\noindent The experiment aims to assess the practical usability of the developed tool \textit{ExtracTable} for creating scientific content for knowledge graphs. To this end, we collaborated with ORKG curators who have previously created comparisons as part of literature reviews. Each participant will first subjectively assess the time they spent on manually creating one of their existing comparisons, reflecting state-of-the-art research in a specific domain. They will then recreate the same comparison using \textit{ExtracTable}, without reviewing the original version to avoid bias. We used a within-subject design to assess usability and enable meaningful self-comparison across diverse domain experts.

The developed framework \textit{ExtracTable} is the independent variable, while effectiveness, efficiency, and satisfaction are the dependent variables for usability. Effectiveness is measured by participants' perception and objectively by the number of identified papers, along with the accuracy of extracted data compared to the established comparison. 
Efficiency is assessed through participants' perception and objectively by the time taken to create the comparison.
Satisfaction is evaluated subjectively after each stage and overall using the System Usability Scale (SUS)~\cite{lewis_item_2018} at the experiment's conclusion.

With particpants' consent, both screen and audio were recorded during the sessions to enable detailed post-analysis. The experiment is described in a document~\cite{John.2025} provided to participants at the start, outlining the study context and presenting title and description of their most recent comparison. \href{https://gitlab.com/TIBHannover/orkg/ExtracTable}{\textcolor{blue}{ExtracTable}} runs locally on a computer provided by the experimenter. Sessions are conducted either in person or remotely via TeamViewer, ensuring a consistent experience. A \href{https://survey.uni-hannover.de/index.php/428288?lang=en}{\textcolor{blue}{questionnaire}} collects demographic data, ORKG usage frequency, and consent. Participants rate their agreement with statements regarding the tools' usefulness and user experience using a Likert scale. The SUS is administered at the end of each session, alongside open-ended questions for additional qualitative feedback.

The target group consists of individuals familiar with the ORKG who have created at least one comparison. 
Participants were recruited through an open call and convenience sampling, with participation being voluntary. 
Each session lasts approximately 60 minutes and follows a structured procedure:
\begin{enumerate}
  \item Introduction to the workflow, questionnaire, and experiment document.
  \item 5-10 minutes for participants to read the provided document.
  \item Completion of initial questionnaire pages, collecting demographic information and baseline data.
  \item Participants proceed through the two stages of \textit{ExtracTable}, responding to related questionnaire items.
  \item Completion of the SUS and provision of general feedback.
\vspace{-0.4cm}
\end{enumerate}

\subsubsection{Results.\label{subsec:results}}
\begin{figure}[!htbp]
  \centering
  \vspace{-0.4cm}
    {
    \includegraphics[width=1\textwidth]{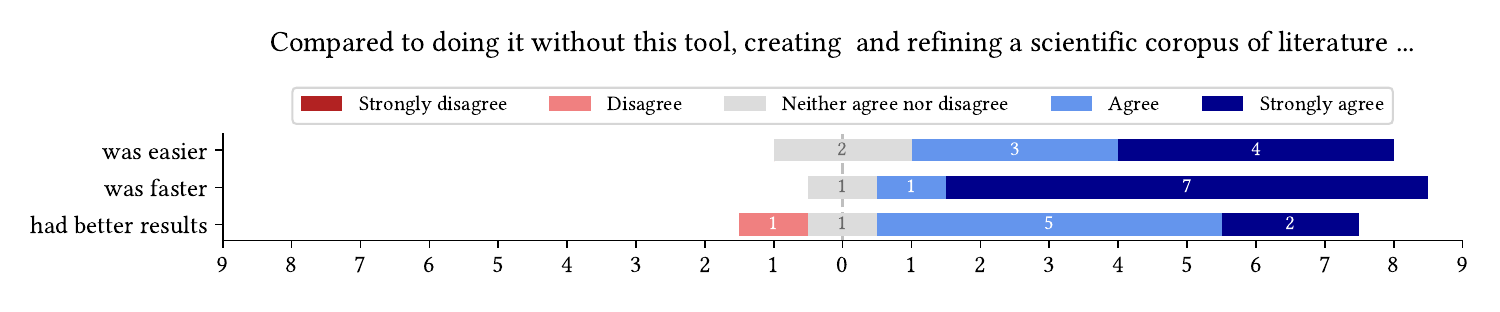}
    \includegraphics[width=1\textwidth]{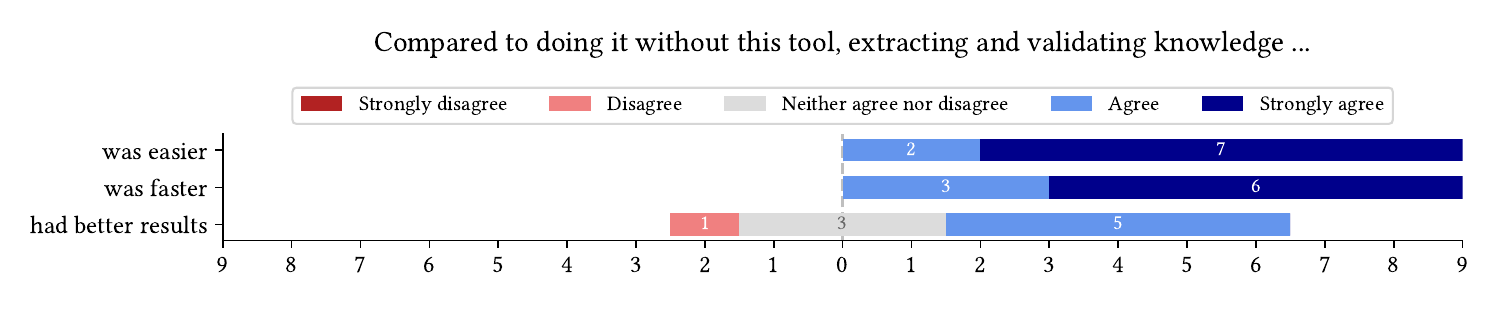}
    
    }
  \caption{Participants' agreement to statements for Stage 1 (above) and 2 (below).}
    \vspace{-0.4cm}
  \label{fig:result_stage_2}
\end{figure}

The raw data~\cite{John.2025} is published on Zenodo. Nine participants took part, including external curators and internal ORKG team members not involved in the tool’s development, mostly from Semantic Web or neuro-symbolic AI fields. Participants reported varied ORKG usage frequencies, ranging from weekly to monthly, with experience spanning from a few months to several years, reflecting different levels of expertise with the system. 
\secref{fig:result_stage_2} presents the participants' responses regarding subjective efficiency and effectiveness, showing strong agreement that the tool makes corpus creation and knowledge extraction `faster' and `easier', and a weaker agreement to its ability to render `better results'.

\begin{figure}[htb]
\vspace{-0.4cm}
    \centering
    \fontsize{6}{8}        
    \includegraphics[width=\textwidth]{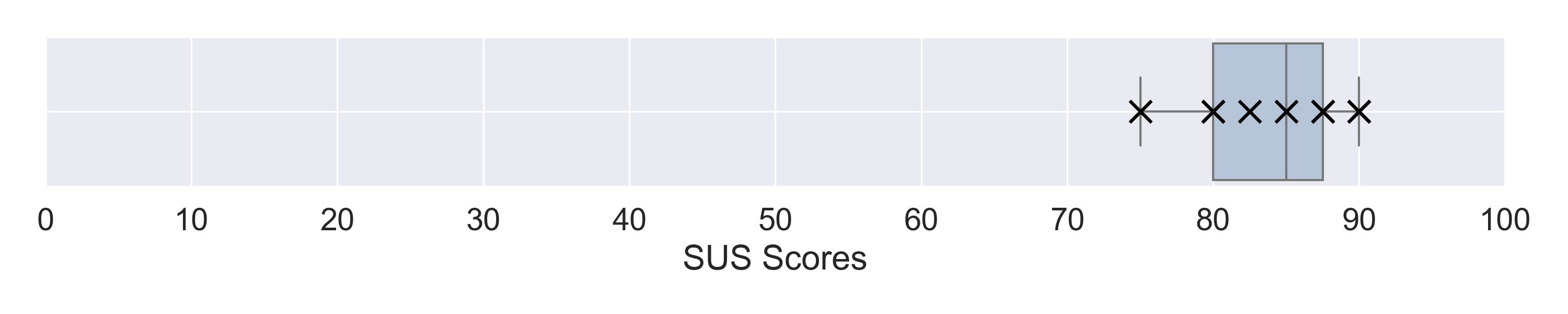}
    \vspace{-0.5cm}
    \caption{SUS scores and boxplot distribution, scoring an average of 84.17 (A+ according to the Lewis and Sauro benchmarks~\cite{lewis_item_2018}).}
    \vspace{-0.4cm}
    \label{fig:sus}
\end{figure}

~\secref{fig:sus} represents the resulting SUS scores.
We achieve an overall SUS score of 84.17 ($min = 75$, $max = 90$, $\sigma = 4.86$), regarded as A+ score -- the highest usability rating~\cite{lewis_item_2018}. Participants expressed general satisfaction with \textit{ExtracTable}, with comments like \textit{``The workflow should be integrated to ORKG as soon as possible. I will like to use it''} and \textit{``Very helpful system!!!''}. Specific feedback highlighted features like the \textit{``Keywords extraction option is an amazing idea''}. Suggestions for improvement included adding \textit{``Properties should have descriptions (for more accurate extraction)''} and \textit{``focus could be on extracting entities that align with the similar papers already available in the ORKG''}.
 
\begin{figure}[!htbp]
\vspace{-0.4cm}
    \centering

    \fontsize{6}{8}
    \includegraphics[width=\textwidth]{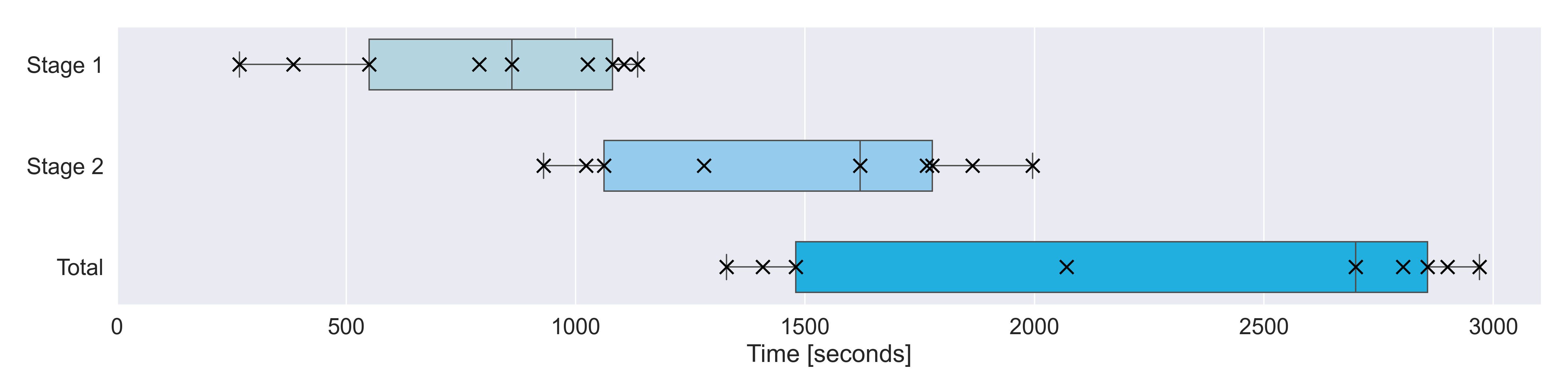}
    \vspace{-0.4cm}
    \caption{Visualization of the time spent by participants across two workflow stages.}
    \label{fig:stages_duration}
    \vspace{-0.4cm}
\end{figure}

The reported durations for the participants' original comparison varied significantly, with the fastest comparison taking approximately 4 hours to complete, while the longest required around 2 weeks. We acknowledge that these self-reported durations are prone to imprecision but still provide an indicative measure of the time involved.
Completing the workflow with our framework -- from defining a research interest based on their original comparison to successfully creating a new ORKG comparison -- took an average of 1,480 seconds (24:40 minutes), with a minimum of 1,023 seconds (17:03 minutes) and a maximum of 1,996 seconds (33:16 minutes), as detailed in~\secref{fig:stages_duration}.

We also assess the tool's ability to replicate the original comparison by evaluating the number of publications it identifies against those included in the original set. 
The newly generated comparisons contained an average of 5 papers ($min$ = 1, $max$ = 8, $\sigma$ = 2.2), whereas the original comparisons had an average of 11 papers. Overall, 98\% of the newly created comparisons differed from the originals, with only 2\% showing overlapping results.
Since the majority of the retrieved papers did not match the original set, it was difficult to objectively assess the quality of the information generated by the tool.\vspace{0.2cm}

\noindent
\textit{QIP.} The QIP methodology showed that curators are a valuable user group for guiding tool development. To improve CSV-based integration of literature tables, enhanced semantification is currently being developed to better align entities within the ORKG~\cite{John_Scimantify.2025}.\vspace{-0.4cm}

\subsubsection{Threats to Validity.}
Based on Wohlin et al.~\cite{wohlin_experimentation_2012}, we identify potential threats to the validity of our experiment.
\textit{Conclusion Validity.} Participants may subconsciously replicate elements from their original ORKG comparison. To reduce this risk, we did not inform them that the task was based on their previous work. Participants came from diverse research fields but worked on their own prior content, minimizing the impact of disciplinary differences while supporting broader generalizability.
\textit{Internal Validity.} Participants may have been predisposed to emphasize the tool’s benefits in comparison to doing it without the tool, particularly ease and speed. As part of the participants were ORKG curators, their financial affiliation could have introduced a bias toward favorable feedback, given the tool's relevance to their ongoing work. Additionally, self-reported durations for prior comparison creation, collected in unobserved settings, are prone to imprecision, adding uncertainty when comparing to the objectively timed tool-based workflow.
\textit{Construct Validity.} To capture genuine reactions and difficulties, we encouraged participants to comment on their experience during each task. This ensured feedback reflected their actual interactions with the tool.
\textit{External Validity.} With only nine participants, generalizability is limited. Nonetheless, the range of disciplinary backgrounds offers insight into the tool’s broader applicability. Future evaluations should involve more diverse and numerous participants to strengthen external relevance.

\section{Discussion and Future Work\label{sec:discussion}}

\noindent
\textit{Experiment.} We interpret the mean SUS score using Bangor et al.'s~\cite{bangor_determining_2009} adjective scale, where \textit{ExtracTable}'s score of 84.17 ranks at the upper boundary of ``Good'', and an A+ in the Lewis and Sauro's benchmarks~\cite{lewis_item_2018}, indicating a positive user experience. Participants described \textit{ExtracTable} as intuitive, easy to navigate, and effective for task execution.
\textit{ExtracTable} significantly reduced the time needed to create content in the ORKG compared to participants' prior methods. 
While this shows strong potential for streamlining workflows, it does not automatically ensure higher-quality outputs.
Participants appreciated the simplified corpus creation and extraction process but expressed concerns about accuracy and over-reliance on automated suggestions.
Those who invested more time refining the corpus reported better outcomes, suggesting deeper engagement improves quality.
Still, not all agreed the tool consistently produced better results, exposing a gap between automation and human expectations. 
This disparity points to a key challenge in HITL systems: balancing automation with human judgment to ensure both efficiency and high-quality, contextually relevant results.
Addressing this requires enhancing the extraction process and incorporating iterative feedback loops to refine accuracy.
Although the experiment aimed to replicate a prior comparison, measuring replication quality through the number of identified papers proved unfeasible as an indicator of replication quality, underscoring the challenge of replicating structured research output in a HITL system.
Unlike automated systems, our approach relies on user insights and interpretations, making the process subjective and context-dependent.\vspace{0.2cm} 

\noindent
\textit{Requirements.} 
\textit{ExtracTable} satisfies most core requirements defined for the HITL framework. It guides users from defining a research interest to extracting and editing data, with CSV export for seamless integration into scholarly KGs like the ORKG. The modular pipeline supports extensibility, adaptability, and long-term semantic preservation.
However, key limitations remain. Entity linking (FR12) is not yet implemented, leaving extracted entities disconnected from existing KG resources. Incorporating entity alignment and exporting in formats like JSON-LD or RDF would improve semantic interoperability.
Additionally, performance degrades with large datasets (NR4), as processing occurs sequentially per document. Future work should focus on performance optimization and parallelization strategies.
Participants also requested more control over the data model (FR8). Allowing users to define domain-specific properties and extraction instructions would improve precision and adaptability across fields. This aligns with feedback on the need for more customizable and explainable outputs.
Crucially, while automation accelerates literature processing, LLM-based extractions still require human validation (FR10, FR11). Without critical oversight, there’s a risk of misinterpreting or blindly trusting generated results. This underscores the ethical implications of AI-assisted research, especially risking the loss of genuine understanding and knowledge internalization.\vspace{0.2cm}

\noindent
\textit{Future Work.}
Future development should allow storing and refining outputs with customizable data models, enabling users to iteratively improve extraction accuracy. Users can selectively re-run extractions on specific cells while preserving high-quality results, enhancing the HITL process. Semantic structuring, like linking content to knowledge graph entities and supporting interoperable formats, will strengthen the system’s utility. Ongoing improvements in scalability and model configurability will stabilize and enrich the tool. Lastly, research should focus on user guidance for validating and interpreting extractions, e.g., by highlighting source passages in PDFs, ensuring human oversight in AI-driven workflows.

\section{Conclusion\label{sec:conclusion}}
Our work contributes \textit{ExtracTable}, a Human-in-the-Loop (HITL) approach for structuring scientific content, demonstrated in the context of the ORKG to showcase its suitability for downstream knowledge graph integration. Aligned with the Quality Improvement (QIP) paradigm, \textit{ExtracTable} comprises a two-stage workflow — \textit{Creating a scientific corpus of literature} and \textit{Knowledge extraction and validation} — implemented as a modular framework in a Jupyter notebook environment. This framework assists researchers in identifying relevant publications, extracting key information, and structuring knowledge into reusable formats ready for integration into scientific knowledge graphs.

By automating repetitive tasks and enabling researchers to interactively refine outputs, \textit{ExtracTable} reduces the manual effort typically involved in processing large volumes of literature. It moves toward making large-scale literature reviews more efficient and helps researchers keep pace with the rapidly expanding body of scientific knowledge. While evaluation results showed the tool is highly usable and accelerates the process, we identified areas for improvement. Additional support is needed to engage users in the knowledge extraction process, alongside feedback loops to further enhance the quality of extracted information.

\section*{Acknowledgements}
The authors would like to thank the Federal Government and the Heads of Government of the Länder, as well as the Joint Science Conference (GWK), for their funding and support within the framework of the NFDI4Ing and NFDI4DataScience consortia. This work was partially funded by the German Research Foundation (DFG) - project number 442146713 and project number 460234259.

\bibliographystyle{splncs04}
\bibliography{Paper}
\end{document}